\documentclass{emulateapj}
\usepackage{apjfonts}
\usepackage{epsf}
\begin{document}

\slugcomment{Submitted to ApJ}
\shortauthors{J. M. Miller et al.}
\shorttitle{NuSTAR and GRS 1915$+$105}

\title{NuSTAR Spectroscopy of GRS 1915$+$105: Disk Reflection, Spin, and Connections to Jets}

\author{J.~M.~Miller\altaffilmark{1}, M. L. Parker\altaffilmark{2},
  F. Fuerst\altaffilmark{3}, M. Bachetti\altaffilmark{4,5},
  F. A. Harrison\altaffilmark{3}, D. Barret\altaffilmark{5},
  S. E. Boggs\altaffilmark{6}, D. Chakrabarty\altaffilmark{7},
  F. E. Christensen\altaffilmark{8}, W. W. Craig\altaffilmark{9,10},
  A. C. Fabian\altaffilmark{2}, B. W. Grefenstette\altaffilmark{3},
  C. J. Hailey\altaffilmark{10}, A. L. King\altaffilmark{1},
  D. K. Stern\altaffilmark{11}, J. A. Tomsick\altaffilmark{6},
  D. J. Walton\altaffilmark{3}, W. W. Zhang\altaffilmark{12}}

\altaffiltext{1}{Department of Astronomy, The University of Michigan, 500
Church Street, Ann Arbor, MI 48109-1046, jonmm@umich.edu}

\altaffiltext{2}{Institute of Astronomy, The University of Cambridge,
  Madingley Road, Cambridge, CB3 OHA, UK}

\altaffiltext{3}{Cahill Center for Astronomy and Astrophysics,
  California Institute of Technology, Pasadena, CA, 91125}

\altaffiltext{4}{Universite de Toulouse, UPS-OMP, Touluse, France}

\altaffiltext{5}{CNRS, Institut de Recherche en Astrophyique et
  Planetologie, 9 Av. colonel Roche, BP 44346, F-31028, Toulouse cedex
  4, France}

\altaffiltext{6}{Space Sciences Laboratory, University of California,
  Berkeley, 7 Gauss Way, Berkeley, CA 94720-7450}

\altaffiltext{7}{Kavli Institute for Astrophysics and Space Research,
  Massachusetts Institute of Technology, 70 Vassar Street, Cambridge,
  MA 02139}

\altaffiltext{8}{Danish Technical University, Lyngby, DK}

\altaffiltext{9}{Lawrence Livermore National Laboratory, Livermore, CA}

\altaffiltext{10}{Columbia University, New York, NY 10027, USA}

\altaffiltext{11}{Jet Propulsion Laboratory, California Institute of
  Technology, 4800 Oak Grove Drive, Pasadena, CA 91109}

\altaffiltext{12}{NASA Goddard Space Flight Center, Greenbelt, MD 20771}


\begin{abstract}
We report on the results of spectral fits made to a {\em NuSTAR}
observation of the black hole GRS~1915$+$105 in a ``plateau'' state.
This state is of special interest because it is similar to the
``low/hard'' state seen in other black holes, especially in that
compact, steady jets are launched in this phase.  The 3--79~keV
bandpass of {\em NuSTAR}, and its ability to obtain
moderate--resolution spectra free from distortions such as photon
pile-up, are extremely well suited to studies of disk reflection in
X-ray binaries.  In only 15~ks of net exposure, an extraordinarily
sensitive spectrum of GRS 1915$+$105 was measured across the full
bandpass.  Ionized reflection from a disk around a rapidly--spinning
black hole is clearly required to fit the spectra; even hybrid
Comptonization models including ionized reflection from a disk around
a Schwarzschild black hole proved inadequate.  A spin parameter of $a
= 0.98\pm 0.01$ ($1\sigma$ statistical error) is measured via the
best--fit model; low spins are ruled out at a high level of confidence.
This result suggests that jets can be launched from a
disk extending to the innermost stable circular orbit.  A very steep
inner disk emissivity profile is also measured, consistent with models
of compact coronae above Kerr black holes.  These results support
an emerging association between the hard X-ray corona and the base of
the relativistic jet.
\end{abstract}

\section{Introduction}
Reflection of hard X-ray emission from a ``corona'' onto the accretion
disk can measure black hole spin, and can also serve as a powerful
probe of the geometry of black hole accretion flows.  Disk reflection
spectra excited near to black holes will bear the imprints of
gravitational red-shifts and strong Doppler shifts (e.g. Fabian et
al.\ 1989).  As long as the accretion disk extends to the innermost
stable circular orbit (ISCO; Bardeen, Press, \& Teukolsky 1972), the
degree of the distortions imposed by these shifts can be used to infer
the spin of the black hole; efforts to exploit disk reflection as a spin
diagnostic in X-ray binaries began in earnest over a decade ago.
Owing to the fact that the effects on Fe K emission lines are
especially pronounced features, and owing to the high flux levels
observed in Galactic X-ray binaries, spin measurements have been made
in a number of systems using this technique (e.g. Miller 2007; Miller
et al.\ 2009).

In cases where the disk extends to the ISCO and the continuum is known
to be fairly simple, not only can spin be inferred, the geometry of
the corona can also be discerned.  The best spectra and variability
studies appear to point toward a very compact central corona ($r
\leq$10--20~ $GM/c^{2}$; e.g. Reis \& Miller 2013), consistent with
prior results suggesting that hard X-ray emission may arise in the
base of a relativistic jet (e.g. Fender et al.\ 1999; Markoff, Nowak,
\& Wilms 2005; Miller et al.\ 2012).  However, this is not yet clear,
and it also unclear that this geometry holds universally.    

Extremely high sensitivity -- especially over a broad spectral band --
provides a path forward in situations where the continuum and
reflection spectrum may be more difficult to parse.  {\em NuSTAR} detectors
have a triggered read-out; unlike CCD spectrometers, they are not
subject to pile-up distortions (Harrison et al.\ 2013).  In this
respect, {\em NuSTAR} is especially well-suited to disk reflection studies
of bright Galactic compact objects.  Moreover, {\em NuSTAR} offers
unprecedented sensitivity out to almost 80~keV, giving an excellent
view of the Compton back-scattering hump (typically peaking in the
20--30~keV), and any additional curvature or breaks.

GRS 1915$+$105 is a particularly important source for understanding
black hole spin, disk--jet connections in all accreting systems, and
how accretion flows evolve with the mass accretion rate.  Prior
efforts to measure the spin of GRS 1915$+$105 have not come to a clear
consensus.  Moreover, a multiplicity of states are observed in GRS
1915$+$105 (Belloni et al.\ 2000).  The most intriguing of these may
be the so-called ``plateau'' state, because it bears the closest
analogy with the ``low/hard'' state in other black hole transients.
Notably, radio emission consistent with compact jet production and
strong low--frequency quasi-periodic oscillations (QPOs) are observed
in this state (e.g. Muno et al.\ 2001); when combined with sensitive
spectroscopy, these features may offer unique insights into the inner
accretion flow.

In Section 2, we describe the {\em NuSTAR} observation of GRS 1915$+$105 and
our reduction of the data.  Section 3 describes our analysis of the
FPMA and FPMB specta.  In Section 4, we discuss the results of our
spectral fits and their impacts.

\section{Observations and Data Reduction}
{\em NuSTAR} observed GRS 1915$+$105 on 03 July 2012, over a span of
59.8~ks.  The data were screened and processed using NuSTARDAS version
1.1.1.  Spectra from the FPMA and FPMB detectors were extracted from
90'' regions centered on the source position.  Background spectra were
extracted from regions of equivalent size on each detector; however,
the background is negligible.
Response files appropriate for the pointing (on-axis), source type
(point, not extended) and region size were automatically created by
the NuSTARDAS software.  After all efficiencies and screening, the net
exposure time for the resultant spectra was 14.7~ks for the FPMA, and
15.2~ksec for the FPMB.  The net observing time is small compared to
the total observing due to the source flux, and in part because the
observation occurred very early in the mission, and in part owing to
detector dead-time.

The spectra were analyzed using XSPEC version 12.6 (Arnaud \& Dorman
2000).  The $\chi^{2}$ statistic was used to assess the relative
quality of different spectral models.  We used ``Churazov'' weighting
for all fits to govern the influence of bins with progressively less
signal at high energy (Churazov et al.\ 1996).  All errors reported in
this work reflect the $1\sigma$ confidence interval on a given
parameter.

\section{Analysis and Results}
Examination of the {\it Swift}/BAT light curve of GRS~1915$+$105 shows
that our observation was made at the start of an $\sim$100-day
interval with sustained hard flux and only moderate variability.
Intervals before and after have much stronger day-to-day variability.
The lightcurve of our observation shows significant source variability
on short time scales, typical of GRS 1915$+$105, as well as moderately
strong QPOs between 0.5--3.0~Hz.  A full timing analysis will be
reported in a separate paper (Bachetti et al.\ 2013, in preparation),
but the fact of these variability properties helps us to make a secure
identification of the source state.  These timing properties, as well
as the source flux observed by the {\it Swift}/BAT, are typical of the
``plateau'' state of GRS 1915$+$105 (e.g., Muno et al.\ 2001,
Trudolyubov 2001, Fender \& Belloni 2004).  Observations with the
RATAN-600 radio telescope found that GRS 1915$+$105 varied between $12\pm3$
and $6\pm3$ mJy at 4.8 GHz (Trushkin, private communication) during
the {\em NuSTAR} observation, consistent with relatively radio--faint
``plateau'' states.

Version 1.1.1 of the NuSTARDAS software and calibration has verified the
detector response over the 3--79~keV band, partly through careful
comparisons to the Crab.  In all cases, the FPMA and FPMB spectra of
GRS 1915$+$105 were jointly fit over the 3--79~keV band.  An overall
constant was allowed to float between the detectors to account for any
mismatch in their absolute flux calibration; in all cases, the value
of this constant was found to be 1.02 or less.  In all fits,
absorption in the ISM was fit using the ``tbabs'' model (Wilms, Allen,
\& McCray 2000), using corresponding abundances (``wilm'') and cross
sections (``vern''; Verner et al.\ 1996).

Figure 1 shows the FPMA and FPMB spectra of GRS 1915$+$105, fit with a
basic power-law model.  The sensitivity of both spectra is excellent.
Simple, broken, and cut-off power-law models all fail to fit the data.
However, they approximate the continuum, and the prominence of the
remaining disk reflection features in the spectra is readily discerned
in the ratio plots in Figure 2.  The power-law indices obtained in
these simple fits are broadly consistent with values measured in fits
to {\it Suzaku} spectra of GRS 1915$+$105 in the ``plateau'' state (Blum et al.\ 2009).

The ``comptt'' model describes thermal Comptonization (Titarchuk
1994).  It also leaves strong reflection-like residuals, and
does not provide an acceptable fit.  The ``nthcomp'' model is
essentially a more physical means of obtaining a cut-off power-law
continuum by mixing thermal and non-thermal electron distributions
 (Zycki, Done, \& Smith 1999).
Importantly, ``nthcomp'' is capable of accounting for curvature that
might otherwise be mistaken for disk reflection.  However, the
data/model ratio and fit statistic in Figure 2 show that even
``nthcomp'' is unable to account for the strong, broad Fe K line and
the Compton back-scattering hump.

The ``eqpair'' model describes Compton scattering in a
sophisticated way, allowing mixtures of thermal and
non-thermal electron distributions (Coppi
1999). ``Eqpair'' also explicitly includes blurred disk reflection.
However, the reflection spectrum (described via the ``pexriv'' model,
Magdziarz \& Zdziarski 1995) is blurred with the ``rdblur'' function
(Fabian et al.\ 1989), which only describes the Schwarzschild metric
and does not permit spin measurements.  The internal reflection was
coupled to an external ``diskline'' model, which is the kernel of
``rdblur'', in order to account for the emission line.  In our fits,
we fixed the cosine of the inclination angle to 0.3, the elemental
abundances to solar values, the inner disk radius to minimum possible
$r_{in} = 6~GM/c^{2}$, the outer radius to $r_{out} = 1000~ GM/c^{2}$,
the emissivity to the Euclidian value of $q = 3$ (recall that $J
\propto r^{-q}$) and the disk temperature to $T = 10^{6}$~K (the
maximum allowed).  The reflection fraction and disk ionization were
allowed to vary.  Numerous parameters control the hybrid thermal and
non-thermal continuum.  For simplicity, we fixed the disk blackbody
temperature from which photons are up-scattered to $kT = 0.2$~keV, and
varied the soft photon compactness ($l_{bb}$), the ratio of the hard
to soft compactness ($l_{h}/l_{s}$), the fraction of the power
supplied to energetic particles that goes into accelerating
non-thermal particles ($l_{nt}/l_{h}$), and the Thomson scattering
depth ($\tau$).  The radius of the scattering region could not be
constrained and was fixed at $1.5\times 10^{6}~ {\rm cm}$.  Default
values were assumed for all other parameters.  Fitting ``eqpair'' in
this way, a large improvement is achieved ($\chi^{2}/\nu = 5529/3784$
(see Figure 2).  

Given these results, models focused on ionized disk reflection were
next pursued.  Our best-fit spectral model is ${\rm constant} \times
{\rm tbabs} \times (({\rm kerrconv}\times {\rm reflionx\_hc}) + {\rm
  cutoffpl})$ (see Table 1, and Figures 3 and 4). ``Kerrconv'' is a
relativistic blurring function, based on
ray-tracing simulations (Brenneman \& Reynolds 2006).  It includes
inner and outer disk emissivity indices (following Wilkins \& Fabian
2012, $q_{1}$ floated freely but $q_{2} \geq 0$ was required), an
emissivity break radius, the black hole spin parameter, the inner disk
inclination (bounded between $65^{\circ} < i < 80^{\circ}$, based on
jet studies by Fender et al.\ 1999), and inner and outer disk radii
(in units of the ISCO radius; values of $r_{in} = 1.0$ and $r_{out} =
400$ were frozen in all fits).  ``Reflionx\_hc'' is a new version of
the well-known ``reflionx'' model that describes reflection from an
ionized accretion disk of constant density (Ross \& Fabian 2005),
assuming an incident power-law with a cut-off.  These models capture
important effects by solving the ionization balance within the disk,
and scatter-broadening photoelectric absorption edges.  That is,
``reflionx\_hc'' includes broadening due to scattering, and this
effect is balanced against dynamical and gravitational broadening when
the model is convolved with ``kerrconv''.  The power-law index of the
hard emission in the ``reflionx\_hc'' and ``cutoffpl'' models was
linked in our fits, as was the characteristic exponential cut-off
energy.  The abundance of Fe within ``reflionx\_hc'' was allowed to
vary in the $1.0 \leq A_{Fe} \leq 2.0$ range, and the ionization
parameter was allowed to float freely ($\xi = L/nr^{2}$).  Flux
normalizations for the ``reflionx\_hc'' and ``cutoffpl'' models were
also measured.

As shown in Table 1, the best blurred reflection model gives a fit
statistic of $\chi^{2}/\nu = 4070.6/3785$.  This model returns a
precise spin measurement: $a = 0.98\pm 0.01$.  The quoted error is only
the statistical error.  Systematic errors are likely much larger, and
related to the assumption that the optically--thick disk truncates at
the ISCO (see, e.g., Shafee et al.\ 2008; Reynolds \& Fabian 2008;
Noble, Krolik, \& Hawley 2010), and different methods and physics
captured in different spectral models.

To obtain a broader view of the spin measurment and its uncertainty,
we scanned the $0 \leq a \leq 0.998$ range using the ``steppar''
command in XSPEC.  We made an initial scan with 100 points across the
full band, and a second scan with 50 points in the $0.95 \leq a \leq
0.998$ range.  Figure 4 shows the results of this error scan.  There
is a clear minimum at $a = 0.98$; a maximal spin of $a = 0.998$ is
rejected at very high confidence, and so too are low spin values.  It
is notable that the $\chi^{2}$ versus $a$ contour shows local
fluctuations, especially between $0.8 \leq a \leq 0.95$, although all
$\chi^{2}$ values are significantly higher than achieved for the
best-fit spin of $a = 0.98(1)$.  The fluctuations may
indicate deficiencies in the spectral model, or could be partly due to the
limited energy resolution of the spectra.  

We also explored a number of fits with key parameters fixed at
particular values (see Table 1).  The data clearly prefer a solar
abundance of Fe, and the very steep inner emissivity index, for
instance.  The data strongly exclude a model with a much higher
cut-off energy.  As also indicated in Figure 4, the data rule out
reflection from a black hole with zero spin at very high confidence
(the model considered in Table 1 fixes $a=0$ and the emissitivity
indices at $q=3$, appropriate for a ``lamp post'' model in a Schwarzschild
regime).  Importantly, a plausible model for the low/hard state is
also ruled out.  The ``truncated'' model in Table 1 changed the
best-fit model to require an inner disk radius fixed at 20 times the
ISCO, and $q=3$.  We also fit the ``eqpair''
model again, fixing $q=10$; this returned $\chi^{2}/\nu = 8059/3782$,
potentially indicating the importance of spin effects.

The best--fit model in Table 1 gives a flux of $F = 2.07(1)\times
10^{-8}~ {\rm erg}~ {\rm cm}^{-2}~ {\rm s}^{-1}$ (0.1--100~keV).
Adopting the mass and distance values favored by Steeghs et
al.\ (2013), $M_{BH} = 10.1\pm 0.6~ M_{\odot}$ and $d = 11$~kpc, this
flux gives a luminosity of $L = 3.0(5)\times 10^{38}~ {\rm erg}~ {\rm
  s}^{-1}$ (where the error is based on an assumed distance
uncertainty of $\Delta d = \pm 1$~kpc), or an Eddington fraction of
$\lambda = 0.23\pm 0.04$.

\section{Discussion}
We have fit numerous models to an early broad-band {\em NuSTAR} spectrum of
GRS 1915$+$105, obtained in a ``plateau'' state.  The sensitivity of
the spectrum is extraordinary, in that the effects of continuum
curvature and disk reflection can clealy be distinguished.  Models
that predict continuum curvature but which do not include reflection
are unable to provide satisfactory fits.  The data require a
continuum with an exponential cut-off, and reflection from an ionized
accretion disk around a black hole with a spin of $a = 0.98(1)$.

Evidence of a relativistic disk line in GRS 1915$+$105 was first
detected with BeppoSAX (Martocchia et al.\ 2002).  Fits to the line
detected in archival ASCA spectra recorded a steep emissivity and
small inner radius ($r = 1.8~ r_{g}$) commensurate with a spin
approaching $a \simeq 0.9$ (Miller et al.\ 2005; similar values were
subsequently found by McClintock et al.\ 2006 and Middleton et
al.\ 2006 using the disk continuum).  Two observations with XMM-Newton
also detected broad lines but were inconclusive with respect to spin
(Martocchia et al.\ 2006), as was a deep spectrum of GRS 1915$+$105 in
the ``plateau'' obtained with {\it Suzaku} (Blum et al.\ 2009).

The measurement of a high spin parameter in a source known for jet
production is interesting in that it may indicate that spin powers
jet production, as predicted by e.g. Blandford \& Znajek (1977).  It is {\it
  possible} that the jet is powered partly by tapping the spin (Miller
et al.\ 2009; Fender, Gallo, \& Russell 2010; Narayan \& McClintock
2012, Steiner, McClintock, \& Narayan 2013; Russell, Fender, \& Gallo
2013).  However, the broadest survey of available data suggests that
the mass accretion rate and/or magnetic field may act as a kind of
``throttle'' (King et al.\ 2013a, 2013b) and do more to affect jet power.

The spectral fits presented in this paper also offer some potential
insights into the geometry of the inner accretion flow, and into jet
production.  Compared to an Euclidean emissivity of $q=3$, the inner
emissivity index is extremely steep ($q \simeq 10$, see Table 1).
This may ultimately be unphysical or incorrect; however, the same spin
is obtained when $q = 5$ is fixed (see Table 1).  Our results appear
to broadly confirm the predictions of independent ray-tracing studies
that find steep and broken emissivity profiles for compact, on-axis,
hard X-ray sources emitting close to rapidly--spinning black holes
(Wilkins \& Fabian 2011, 2012; Dauser et al.\ 2013).  The emissivity
is also predicted to flatten at moderate radii, again consistent with
our results.  Given that GRS 1915$+$105 launches compact radio jets in
the ``plateau'' state (e.g. Muno et al.\ 2001; Trudolyubov 2001;
Fender \& Belloni 2004), the hard X-ray region may plausibly
associated with the base of the jet.

A very steep inner emissivity profile was recently reported in fits to
the {\it Suzaku} spectrum of Cygnus X-1 in the ``low/hard'' state
(Fabian et al.\ 2012).  Joint {\it Suzaku} and radio monitoring of
Cygnus X-1 in the ``low/hard'' state also concluded that the hard
X-ray continuum is likely produced in the base of the relativistic jet
(Miller et al.\ 2012).  More broadly, similar emissivity profiles have
been seen in massive black holes acceting at relatively high Eddington
fractions, notably 1H 0707$-$495 (Fabian et al.\ 2009).  Studies of
time lags in Seyferts and microlensing in quasars suggest that very
compact coronae may be common (Reis \& Miller 2013).

Advection--dominated accretion flow models predict that the inner disk
should be truncated at $\dot{m}_{Edd} \simeq 0.08$, or $\lambda \simeq
0.008$ (assuming an efficiency of 10\%; Esin, McClintock, \& Narayan
1997).  This is broadly consistent with the luminosity at which many
sources transition into the ``low/hard'' state, wherein jet production
is ubiquitous.  Our results indicate that a steady jet can potentially
be launched from a disk that extends to the ISCO.  The
disk, corona, and jet are undoubtedly a complex, coupled system, but
jet production in black holes may be more closely tied to the nature
of the corona than the inner disk radius.  This may support a new
model for jets and QPOs in accreting black holes (McKinney,
Tchekhovskoy, \& Blandford 2012).

This {\em NuSTAR} observation has offered new insights into nature of the
accretion flow in the ``plateau'' state, owing to its extraordinary
sensitivity.  Similarly, it has also provided the first strong spin
constraint based on disk reflection modeling.  However, additional
modeling using developing disk reflection codes, and a deeper
observation in the ``plateau'' state, are likely required in order to
confirm these initial model--dependent results.  A {\em NuSTAR} observation
in a softer, more luminous state is likely also required in order to
rigidly test and verify the spin measurement.

\hspace{0.1in}

This work was supported under NASA Contract No. NNG08FD60C, and made
use of data from the {\em NuSTAR} mission, a project led by the California
Institute of Technology, managed by the Jet Propulsion Laboratory, and
funded by NASA.  JMM thanks Sergei Trushkin for communicating radio results.

\clearpage

\begin{figure}
\includegraphics[scale=0.5,angle=-90]{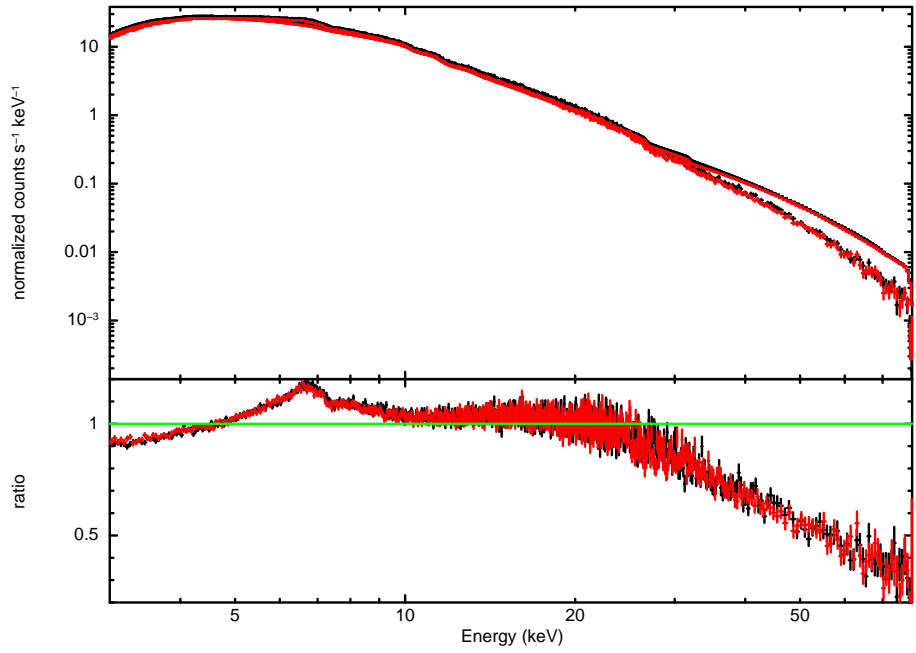}
\figcaption[t]{\footnotesize  The 3--79~keV {\em NuSTAR} FPMA
  (black) and FPMB (red) spectra of GRS 1915$+$105, fit with a simple
  power-law assuming $N_{H} = 6\times 10^{22}~ {\rm cm}^{-2}$.  The
  4.0--8.0~keV and 15.0--45.0~keV bands were ignored in order to
  portray the curvature in the spectrum.  A strong, skewed Fe K line
  is visible in the 4--8~keV band.  The curvature in the 20--30~keV
  band is due to a combination of a spectral cut-off and disk
  reflection.}
\end{figure}
\medskip

\clearpage

\begin{figure}
\includegraphics[scale=0.8]{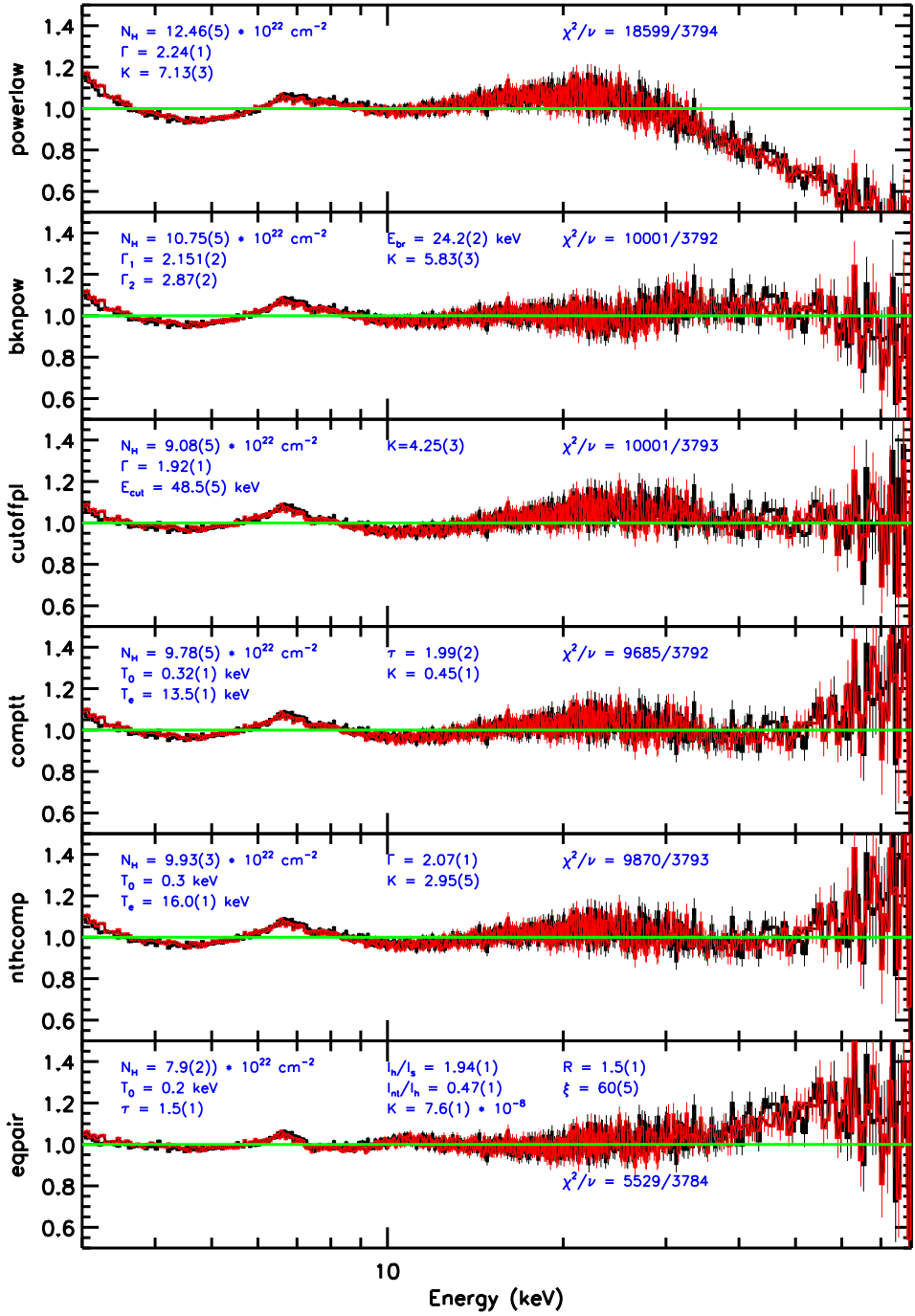}
\figcaption[t]{\footnotesize  The data/model
  ratios obtained when the 3--79~keV FPMA (black) and FPMB (red)
  spectra of GRS 1915$+$105 are jointly fit with common spectral
  models.  In each panel, the name of the spectral model is given on
  the vertical axis.  The key parameters derived from each spectral
  fit, including the $\chi^{2}$ statistic, are given in each panel (in
  XSPEC parlance).  In each case, ``K'' is the flux normalization of
  the model.  Note that even ``nthcomp'' and ``eqpair''
  (sophisticated Comptonization models) fail to describe the spectra
  owing to the strong, blurred reflection features that are present.}
\end{figure}
\medskip

\clearpage

\begin{deluxetable}{llllllllllllll}
\tablecaption{Relativistically--blurred Disk Reflection Models}
\tablefontsize{\tiny}
\tablehead{
\colhead{model} & \colhead{$N_{H}$}  &  \colhead{$q_{in}$}  &  \colhead{$q_{out}$}   &  \colhead{$r_{break}$} &  \colhead{$a$}  & \colhead{$\theta$}  &  \colhead{$\Gamma$} &   \colhead{$E_{cut}$} &  \colhead{$K_{pow}$}  &  \colhead{$\xi$}   &   \colhead{$A_{Fe}$} &   \colhead{$K_{refl}$} & \colhead{$\chi^{2}/\nu$} \cr 
\colhead{}   &  \colhead{($10^{22}~ {\rm cm}^{-2}$)} & \colhead{}  &  \colhead{}  &  \colhead{($r_{g}$)} &  \colhead{($cJ/GM^{2}$)} & \colhead{(deg.)} & \colhead{} & (keV) & \colhead{}  & \colhead{(erg~cm~${\rm s}^{-1}$)} & \colhead{} & \colhead{($10^{-5}$)} & \colhead{} }

\startdata

best-fit &  6.11(3) & 9.97(3) & $0.00^{0.01}$ &  6.5(1) &  0.983(3) &  72(1) &    1.720(2) &  35.6(3)  & 2.59(1) &  1020(10) &   1.00(5) &   1.25(7) &  4070.6/3785 \\

\hline

$A_{Fe} = 2$ &  6.09  &  10.0  &   0.00 &   6.48 &   0.985 &    72.3 &   1.70  &   34.8  &   1.66  &   890  &    2.0*  &   1.15  &     4109.0/3786 \\ 

$q_{in} = 5$ &  6.26  &  5.0*  &    0.55  &   11.1 &   0.998 &   65.5 &   1.72   &    35.5  &  2.57  &   1230  &   1.0  &    0.86   &    4216.0/3786 \\

$r_{break} = 3$ &  6.35  &  6.84  &   1.09  &   3.0*  &   0.977 &   71.9  &  1.74  &     35.6  &   2.59   &  1350 &    1.0  &    0.62  &     4260.9/3786 \\

truncation &  6.17  &  3.0*  &    3.0* &   6.0* &   0.98*  &   65.0  &  1.74   &    36.0  &  2.52  &   1520 &    1.0  &    0.58  &     4547.4/3789 \\

$a = 0$ &  6.44 & 3.0* & 3.0* & 6.0* & 0.0* & 65.0 & 1.77 & 40.0 & 2.10 & 5000 & 1.0 & 0.32 & 6113.2/3789 \\

higher $E_{cut}$ &   6.83 &   9.76  &   0.003  &  6.55 &   0.988 &    74.4  &  1.83   &    55*  &   2.81  &   1330 &    1.0   &   1.28  &     6543.9/3786 \\

\enddata

\tablecomments{The parameters obtained for the
  best-fit relativistically--blurred reflection model,
  $tbabs*kerrconv*(reflionx\_hc + cutoffpl)$.  The cut-off power-law
  normalization, $K_{pow}$, has units of ${photons}~ {\rm cm}^{-2}~
  {\rm s}^{-1}~ {\rm keV}^{-1}$ at 1 keV.  Please see the text for
  additional details.  The table also lists the results obtained for
  various models wherein parameters were fixed in order to explore the
  sensitivity of the fit statistic to plausible variations.  Errors
  were only calculated for the best-fit model; the reported errors are
  $1\sigma$ confidence limits.  Parameters marked with an asterisk
  denote those fixed at a particular trial value in the rejected
  models.  In the ``truncation'' model, the inner radius of the disk
  was fixed at $20\times r_{~~ISCO}$.  }
\end{deluxetable}
\medskip

\clearpage

\begin{figure}
\includegraphics[scale=0.5,angle=-90]{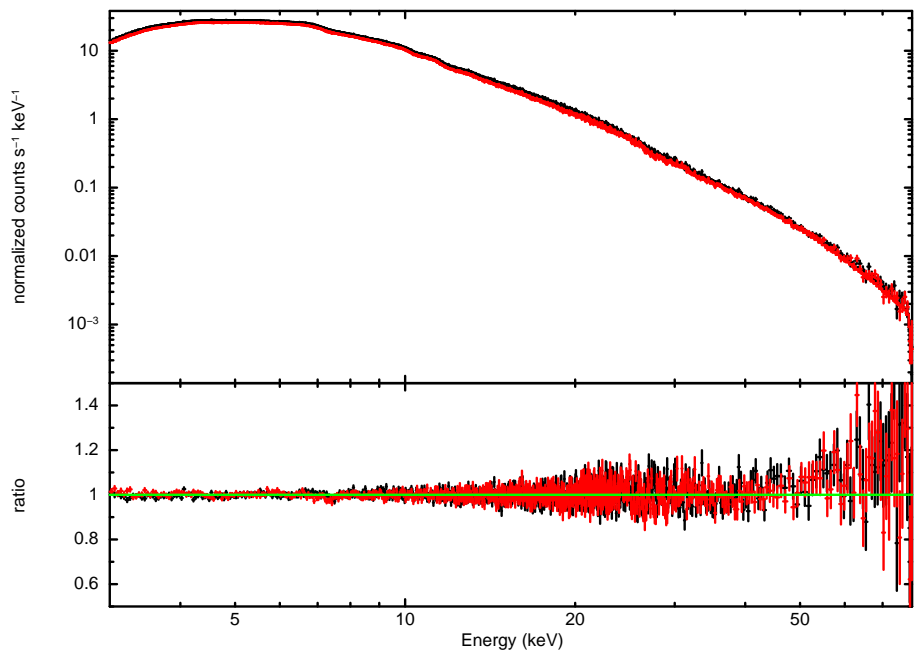}
\figcaption[t]{\footnotesize The FPMA (black)
  and FPMB (red) spectra of GRS 1915$+$105, fit with a
  relativisitically--blurred disk reflection model.  The continuum and
  reflection model include an exponential cut-off, as indicated by the
  simple fits, and consistent with prior results obtained in the
  ``plateau'' state.  Using this model, a black hole spin parameter of
  $a = 0.98(1)$ (statistical error only) is measured (see Table 1).
  The spectra were rebinned for visual clarity.}
\end{figure}
\medskip

\clearpage

\begin{figure}
\includegraphics[scale=0.8]{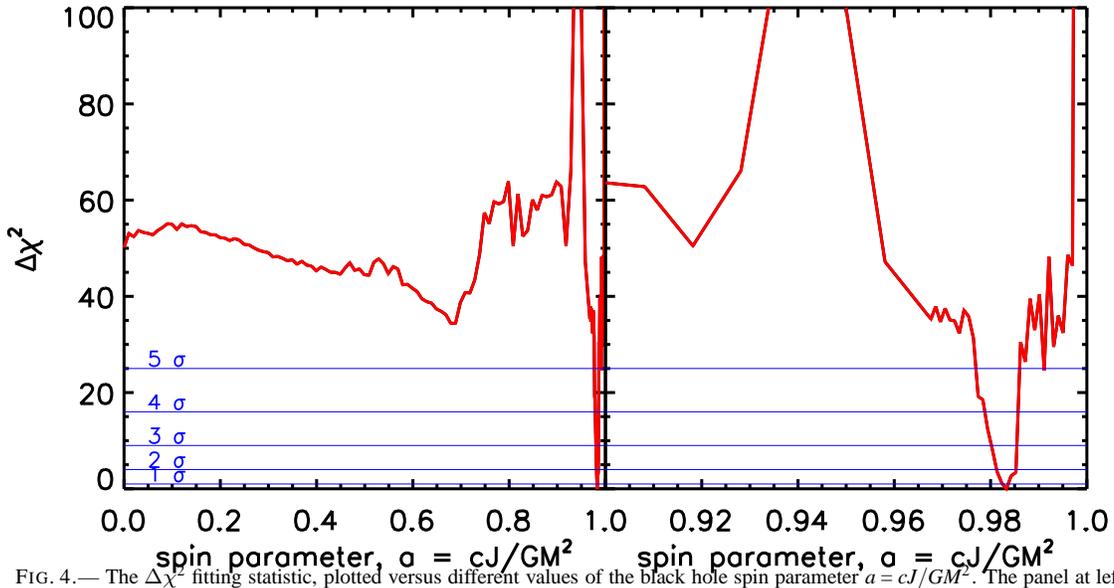}
\figcaption[t]{\footnotesize The $\Delta\chi^{2}$ fitting statistic, plotted versus different values of the
  black hole spin parameter $a = cJ/GM^{2}$.  The panel at left shows
  the full range, while the panel at right shows the $0.9\leq a \leq
  1.0$ range, for clarity.  The spin measurement is based on
  relativistically--blurred disk reflection modeling of the NuSTAR
  spectrum of GRS 1915$+$105 in the ``plateau'' state (see Table 1).
  The error range was scanned using the XSPEC tool ``steppar'', which
  allows all parameters to vary during the scan.  The horizontal
  confidence levels indicate the Gaussian equivalent $\sigma$ value
  for the indicated change in $\chi^{2}$, assuming one interesting
  parameter.}
\end{figure}

\end{document}